\documentclass[fleqn,usenatbib]{mnras}

\usepackage{newtxtext,newtxmath}

\usepackage[T1]{fontenc}

\DeclareRobustCommand{\DE}[3]{#2}
\let\DEthebibliography\thebibliography
\def\thebibliography{\DeclareRobustCommand{\DE}[3]{##3}\DEthebibliography}

\usepackage{subcaption}

\usepackage{graphicx}	
\usepackage{amsmath}	
\defcitealias{Roulston2021}{R21} 

\title[Variable dC stars \& eclipsing CEMP candidate]{An eclipsing CEMP candidate discovered in a search for dwarf carbon stars in post-common envelope binaries}

\author[J.~A.~G.~McLennan]{Jonathan A.~G.~McLennan$^1$,
Jay Farihi$^1$\thanks{E-mail: j.farihi@ucl.ac.uk},
and
Steven G.~Parsons$^2$,
\\
$^1$ Department of Physics and Astronomy, University College London, Gower Street, London WC1E 6BT, UK\\
$^2$ Department of Physics and Astronomy, University of Sheffield, Sheffield, S3 7RH, UK}

\begin{document}

\maketitle

\begin{abstract}
Dwarf carbon stars are dominated by members of the Galactic halo and are thus likely carbon-enhanced metal-poor stars.  In this work, 879 bona fide dwarf carbon stars are characterized by their ground-based light curves, and $p<15$\,d modulation is found to be significant in 31 objects (3.5\,per cent), consistent with starspots and rotation in tidally-locked, post-common envelope binaries.  Among these is an unambiguous halo star that is eclipsing every 1.224\,d, and where the 30\,per cent eclipse depth rules out a white dwarf occulter.  Available {\em Gaia} data do not indicate any tertiary in the eclipsing system, but this remains a possibility and follow-up data are necessary to determine the evolutionary history of this first eclipsing binary among carbon-enhanced stars.  Four of the variable sources exhibit clear multi-year, quasi-sinusoidal trends indicative of magnetic-activity and starspot cycles in rapidly-rotating, dynamo-rejuvenated stars.  These data support a picture where carbon pollution results from wind capture prior to Roche lobe overflow, and the orbital period distribution appears to be moderately shifted to longer periods than carbon-normal, low-mass stars in similar binaries.  The band-combined approach adopted in this work may be more sensitive than prior work using single-bandpass light curves, where at most 19 of 34 binary candidates previously published are independently confirmed here.
\end{abstract}




\section{INTRODUCTION}

Classical carbon stars are identified by strong absorption bands of molecules such as C$_2$ and CN, and on this basis can be inferred to have a carbon-to-oxygen ratio (C/O) exceeding unity. For single stars in the solar mass range, atmospheric C/O $>1$ can only arise during the asymptotic giant branch (AGB), where sufficient carbon may be brought to the surface during the third dredge-up, which reaches the outermost helium-burning layer.

The discovery of G77-61 nearly 50 years ago altered this picture. Identified as a high-proper motion, likely M dwarf with unusually red colors, this dwarf carbon (dC) star prototype exhibits strong C$_2$ and CH bands characteristic of classical carbon stars, but has a parallax placing it firmly on the main-sequence, and unambiguous halo kinematics \citep{Dahn1977}. It remains the only dC star with abundance analysis via high-resolution spectroscopy, and is often included in catalogs of carbon-enhanced metal-poor (CEMP) stars due to its extremely low metallicity [Fe/H] $=-4.0$ \citep{Plez2005}. Owing to its estimated $T_{\rm eff}\approx4100$\,K, its spectrum is dominated by carbon molecular features, and r- and s-process element constraints are insufficient to determine its CEMP class directly, but its position in an A(C) vs.\ [Fe/H] diagram supports a CEMP-no classification \citep{Yoon2016,Arentsen2019}. G77-61 is a well-known binary with a companion consistent with a white dwarf in a 245\,d orbit \citep{Dearborn1986,Whitehouse2023}.

As with the prototype, the bulk of dC stars have Galactic orbits consistent with the ancient halo. Studies of $N\approx1000$ dC stars have demonstrated that 60\,per cent have halo orbits while 30\,per cent are consistent with thick disc kinematics, and thus they are likely to be a metal-poor population, and hence CEMP stars \citep{Farihi2018,Farihi2025}. It is therefore plausible that some fraction formed from carbon-rich material in the early Galaxy as is suspected based on the most metal-poor stars currently known \citep[see e.g.][]{Frebel2015,Hansen2016}. If this is the case, and given main-sequence lifetimes approaching or exceeding a Hubble time, dC stars may represent a critical resource for stellar archaeology.

However, the bulk of known objects are too cool and too faint for detailed abundance analyses via atomic lines and high-resolution spectroscopy in the optical or infrared. At lower spectral resolution where data are available, atmospheric models at the cool end of the main sequence require either carbon-normal or otherwise C/O $<1$ \citep{Gustafsson2008,VanEck2017}. Until both modeling and observations of dC stars are amenable to detailed abundance analysis, it is important to constrain their binary properties to better understand their origin.

There are indications that the overall dC star population has a substantial binary component \citep{Whitehouse2018}.  A modest number of companions that are consistent with white dwarfs, and thus prior mass transfer, have been unambiguously identified via astrometry \citep[three with $p\approx1$--10\,yr;][]{Harris2018}, radial velocity monitoring \citep[ten with $p<10$\,d and two with $p\gtrsim1$\,yr;][]{Margon2018,Roulston2021,Whitehouse2023}, and a few by blue or ultraviolet flux excess \citep{Heber1993,Liebert1994,Green2013,Whitehouse2021}. These binaries represent only a few per cent of the known population, and it is paramount to determine or constrain orbital periods for more stars, and thereby inform evolutionary models and thus the origin of dC stars \citep{Kool1995}.

Photometric surveys cover a large fraction of the sky and provide a means to constrain dC star duplicity for most sources. Via wide-field imaging programs, there are several dozen dC stars that are candidate variables with $p\lesssim10$\,d, and are thus potentially post-common envelope binaries where the dC star is tidally-locked to a dark companion \citep[presumed to be a white dwarf;][]{Whitehouse2021}. This picture is substantiated in several cases where stellar activity indicators such as Balmer line emission are observed in combination with radial velocity changes \citep{Margon2018,Whitehouse2018}. There are 34 candidate dC variables identified within a sample of 944 stars using Zwicky Transient Facility (ZTF) DR5, where only two have radial velocity confirmation, and among the remaining 32 sources, many have low-amplitude variability that awaits confirmation \citep{Roulston2021}. 

This work presents an independent analysis of ZTF data for 1063 dC stars from the Sloan Digital Sky Survey (SDSS). The study aims to take advantage of the additional 5\,yr of data now available via ZTF DR24 as compared to earlier work using DR5 and examines the aforementioned candidates independently using a band-combined light curve analysis. The sample construction, photometry processing, period-search and vetting procedures are described in Sections~2 and 3.  The results of the time-domain analysis are presented in Section~4, including the discovery of the first eclipsing dC star. Section~5 has a discussion and a direct comparison with previous work on dC star light curves in ZTF, and an overview of the most likely evolutionary history. The conclusions are given in Section~6.

\section{SAMPLES AND DATA}\label{sec:methods}

The Zwicky Transient Facility uses the Palomar 48-in Oschin Schmidt Telescope to survey the entire available (northern) sky for transient events, with a cadence of roughly three days and is ideal for short-period photometric variability of relatively faint sources.  The following section outlines the selection of targets, the retrieval of their ZTF photometry, the adopted data quality cuts, the construction of light curves and their analysis.

\subsection{Target samples}

Sample I consists of 1063 bona fide dC stars, distilled from a larger catalog of 1211 spectroscopic candidates \citep{Green2013} so that each source lies on or below the main sequence in three different color-absolute magnitude diagrams \citep{Farihi2025}. Sample II consists of 34 candidate dC stars reported as photometrically variable by \citet[][hereafter R21]{Roulston2021}, identified on the basis of their ZTF DR5 photometry using individual bandpass light curves.  Data for all targets are sourced from the latest ZTF data release, DR24.

\subsection{Sources of potential photometric contamination}

Each dC star has been cross-matched by position with {\em Gaia} DR3 to obtain a unique source ID, right ascension, declination, astrometry, and photometry \citep{Gaia2023}. A cone search is then performed around each {\em Gaia} target with the aim to characterize any neighboring sources that might adversely affect the ZTF photometry and subsequent light-curve analysis.

To designate a light curve as unaffected by source confusion, the acceptance criteria are as follows for neighboring objects with magnitude $m$ and $\Delta m$ relative to the target \citep{Guidry2021}:

\begin{itemize}
    \item $\Delta m\geq2$\,mag within 5.0\,arcsec
    \item $\Delta m\geq1$\,mag within 5.0--7.5\,arcsec
    \item $\Delta m<2$\,mag within 7.5--12\,arcsec
    \item $m\geq13$\,mag within 30\,arcsec
    \item $m\geq10$\,mag within 60\,arcsec
\end{itemize}

\subsection{Data retrieval}

For each dC star, time-series photometry is obtained from ZTF DR24 in the $g$, $r$, and $i$ bands \citep{Bellm2019}. All ZTF detections within a 2\,arcsec cone around each target at the propagated sky position are retrieved at each observation epoch. This procedure yields multiple ZTF object IDs for the same astrophysical source across observing seasons, and all matching light-curve segments are collected and associated with the single {\em Gaia}-identified source.

Poor quality observations are then filtered following the guidelines given in ZTF documentation and the wider literature \citep{Masci2019,Guidry2021}. To be passed on for analysis, all data must meet the following requirements:

\begin{itemize}
    \item Unaffected by known artefacts (CATFLAGS $=0$)
    \item Good observation without blending ($\upchi \leq 4$)
    \item Source brightness $m\leq$ LIMITMAG $-0.5$
    \item Sharpness consistent with a point source (SHARP $\leq 0.5$)
\end{itemize}

After these quality cuts, the per-band observations and corresponding uncertainties for each source are converted from magnitude to flux. Each band is then normalized independently to its mean flux before the bands are concatenated into a single, multi-band light curve.  This band-combined approach is based on the expectation that starspot dimming does not have a strong wavelength dependency across the $gri$ filters.  This assumption is verified by results in Section~4.1, and the band-combined light curves improve temporal sampling.

A light curve is considered usable for variability analysis if, after band concatenation and data filtering, the final combined light curve contains at least 100 data points and the star meets the source confusion criteria. Under all criteria, 879 of the original 1063 dC stars have usable light curves, and form the basis for the analyses of Sample I presented below.

\subsection{Further processing and visual inspection}

Instrumental or astrophysical light-curve drifts, including possible long-timescale changes associated with stellar activity cycles, are fitted with a Gaussian process in time using a squared-exponential kernel \citep{Foreman-Mackey2017,Hippke2019}. The kernel correlation length is initialized at 500\,d and then optimized within broad bounds, so that the model captures smooth secular structure and eschews any short-term variations.  The normalized fluxes are then divided by the inferred trend to flatten the light curve. To prevent outlying data points from biasing subsequent variability searches, a one-time $5\upsigma$ clip is applied to the flattened light curve about its median. The surviving points are then re-normalized to unit mean to produce a cleaned, detrended light curve for analysis.  Additionally, without any detrending, each of the 879 light curves in Sample I is visually inspected to identify long-term modulation due to stellar activity cycles (see Section~4.2).

\section{PERIODOGRAM ANALYSIS}

A Lomb--Scargle periodogram is computed for each source light curve using the implementation in the \textsc{astropy} package, and is well suited to the uneven sampling in ZTF data. For each light curve, frequencies from 0.05 to 24\,d$^{-1}$ are searched using ten samples per independent peak, ensuring oversampling of the power spectrum and reducing the chance of missing narrow peaks in the periodogram. This frequency grid encompasses the rotation and orbital periods to which ZTF is sensitive via photometric variability in tidally-locked binaries.

\subsection{First cut using analytic false-alarm thresholds}

To efficiently assess the 879 light curves for the stars in Sample I, the significance of the highest peak in each Lomb--Scargle periodogram is estimated using the extreme-value approximation of \citet{Baluev2008}. For a peak of Lomb--Scargle power $z$, the false-alarm probability is denoted $X(z)$ and approximated by

\begin{equation}
X(z) \approx 1 - P_1(z)\,e^{-\uptau(z)},
\end{equation}

\medskip
\noindent
where $P_1(z)$ is the single-frequency cumulative distribution function under the null hypothesis, and $\uptau(z)$ accounts for the look-elsewhere effect over the searched frequency range.  For each light curve, a false-alarm probability of 99.9\, per cent, corresponding to $X_\star = 0.001$ is adopted, and the corresponding threshold power $z_\star$ is defined by $X(z_\star)=X_\star$. For convenience, the following definition is adopted

\begin{equation}
S \equiv -\log_{10}[X(z)],
\end{equation}

\medskip
\noindent
so that any peaks with $z>z_\star$ and thus $S>3$ are initially identified as significant variability candidates in this work. This approach is computationally inexpensive for a large number of light curves, yields per-source false-alarm thresholds in milliseconds, and should be relatively accurate for ground-based sampling patterns \citep{Suveges2015}.

\subsection{Identification of spurious periodogram signals}

Time-series photometric surveys carried out by ground-based observatories naturally introduce periodogram peaks near the sidereal day and lunar cycles, and their aliases. To avoid these spurious signals, a window function periodogram is computed for each star by evaluating the Lomb--Scargle power of a unit-flux series using the timestamps of the actual light curve. The window function peaks are then cross-matched with the light curve periodogram using a tolerance of 0.02\,d$^{-1}$, which is sufficiently wide to absorb spectral leakage around sharp features but sufficiently narrow not to over-mask \citep{VanderPlas2018}. Any frequency peak coincident with a window function peak or with an obvious window alias is discarded, which substantially reduces the number of candidate signal peaks.  Furthermore, other narrow frequency ranges are masked based on ZTF source classification efforts \citep[e.g.\ 0.48--0.52\,d$^{-1}$;][]{Coughlin2021}.

Real signals can still appear multiple times in a periodogram because of the sampling pattern (aliasing) or owing to non-sinusoidal variability (harmonics). After masking based on the window function and ZTF recommendations, the remaining peaks are searched for alias groups of the form $f \pm n f_{\rm samp}$ within $\pm0.02$\,d$^{-1}$, where $f_{\rm samp}$ denotes the dominant sampling frequency associated with the observing cadence, and harmonics up to $4f$ for each candidate peak. For any harmonics identified, only the lowest frequency is retained as the candidate fundamental frequency, while the cap at $4f$ prevents over-flagging harmonics when $f$ is small.

\begin{figure*}
\centering
\includegraphics[width=\textwidth]{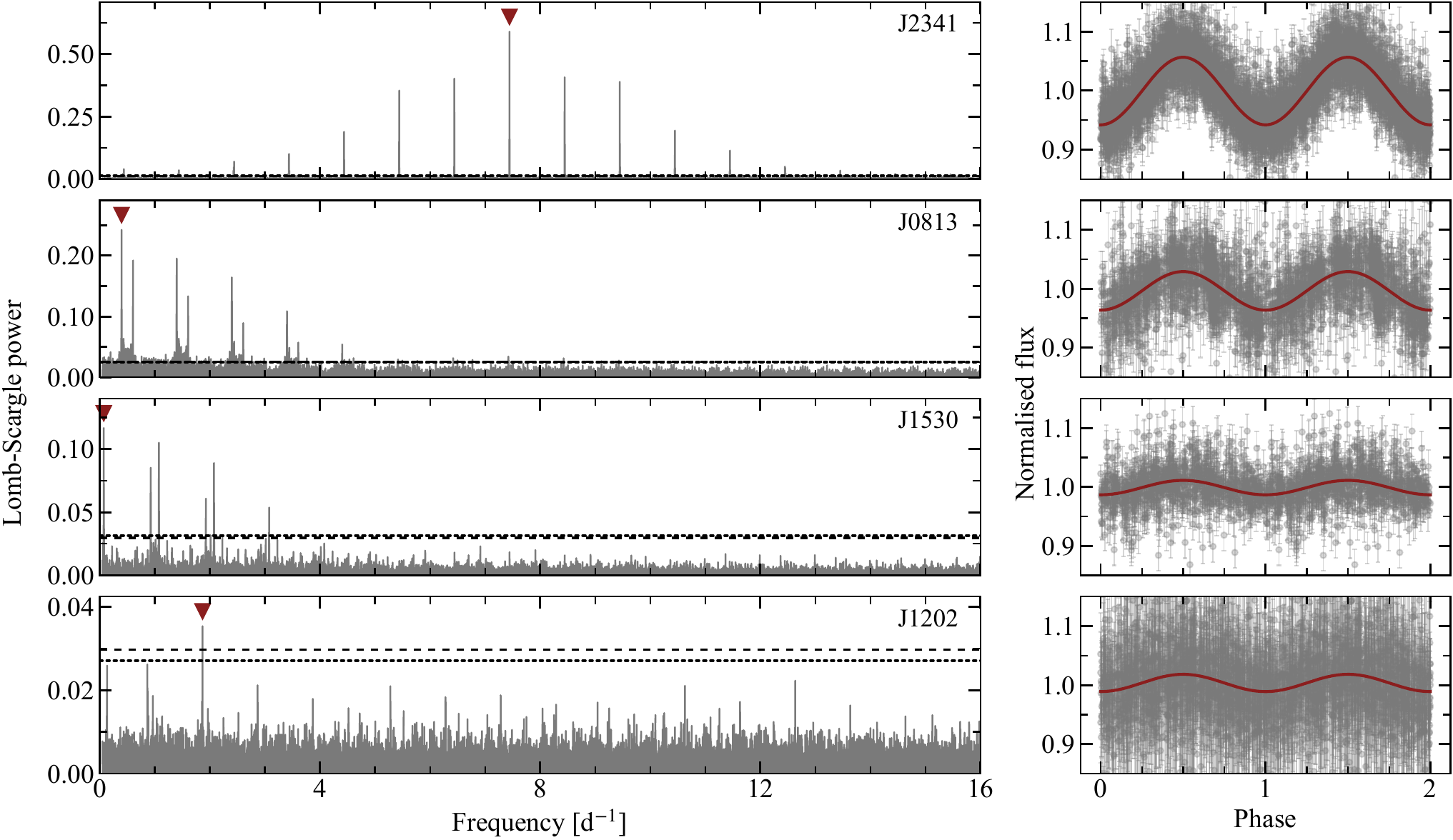}
\caption{Four examples of candidate, photometrically variable dC stars, selected to represent a wide range of signal confidence, and ordered from top to bottom from highest to lowest peak frequency significance. The left-hand panels are periodograms with the adopted frequency highlighted by a red inverted triangle; the dashed and dotted horizontal lines represent the bootstrap and Baluev false-alarm thresholds, respectively. The right-hand panels are the corresponding phase-folded light curves, with phase $=0$ at photometric minimum. The stars are labeled by the first four digits of their J2000 coordinates given in Table~\ref{tab:candidates}.}
 \label{fig:showcase}
\end{figure*}

\subsection{Evaluation of candidate periodogram frequencies}

The surviving candidate frequencies are vetted in two stages. First, the periodograms are visually inspected to ensure that the adopted peak is isolated, narrow, and not embedded in a rising low-frequency noise floor (interpreted as time-correlated noise). Peaks that lie marginally below $z_\star$ in otherwise clean periodograms are earmarked for later bootstrap confirmation using the \textsc{astropy} implementation. Second, the light curve is phase-folded on the candidate frequency, and phase-binned points are overplotted to verify that the modulation is coherent and persists across different parts of the time series. Only those signals that pass both checks are retained for further interpretation.

The remaining periodogram peak with the highest power is adopted for each source as the working frequency for subsequent analysis, while any obvious alias or harmonic family members are noted. For the 31 sources in Sample I that pass the above criteria, and for the stars in Sample II which pass the ZTF data quality requirements detailed in Section~2, more rigorous analyses are performed.  A bootstrap threshold Lomb--Scargle power is calculated for each light curve by re-computing its periodogram 10\,000 times with the observing dates randomly shuffled for each iteration.  The minimum of the resulting, highest ten periodogram peaks then provides a power threshold corresponding to a 99.9\,per cent false-alarm probability.  Each signal that is deemed significant using the bootstrap false-alarm threshold is analyzed using {\sc period04} \citep{Lenz2005}.  Using this tool, the most significant signal frequencies and their amplitudes are refined, and assigned uncertainties based on Monte Carlo simulations.

Despite these several vetting processes, in many cases the aforementioned methodology cannot conclusively identify the fundamental frequency of a true variable source based on photometry alone.  Additional indicators such as radial velocity changes, verification with independent telescopes and instruments, or stellar activity diagnostics are necessary to confirm a genuine rotation or orbital motion when strong window function power may overlap the signal \citep{VanderPlas2018}.

\section{RESULTS}

This section details the outcome of the variability search for the dC star Samples I and II, including the first discovery of an eclipsing dC star, which is unexpectedly consistent with an unevolved companion.

\subsection{Candidate variability detections}

Table~\ref{tab:candidates} lists all the candidate, variable dC stars that pass the extreme-value approximation and bootstrap false-alarm thresholds in their respective periodograms, as well as the aforementioned additional appraisals of the data.  In Figure~\ref{fig:showcase} are four Sample I targets that span a wide range of signal peak significances for their most likely frequencies. For each object, the Lomb--Scargle periodogram is shown with the adopted frequency, both the extreme-value approximation and bootstrap false-alarm thresholds, and the corresponding phase-folded light curve. The top periodogram has the highest significance and is unambiguous, and the phase-folded light curve exhibits a coherent, low-scatter waveform, indicating robust and statistically significant periodic variability.  In contrast, the periodogram at the bottom has a candidate frequency that is closer to the calculated false-alarm thresholds.  In this latter case, the adopted frequency peak still passes all selection criteria, but the phase-folded light curve exhibits noticeably larger scatter, yielding comparatively weaker evidence for coherent modulation. Most candidates are therefore not treated as secure periodic variables based on photometric evidence alone, and require independent confirmation, such as radial-velocity monitoring or higher-cadence time-series photometry.

\subsection{Long-term light curve variations}

\begin{figure}
\centering
\includegraphics[width=\columnwidth]{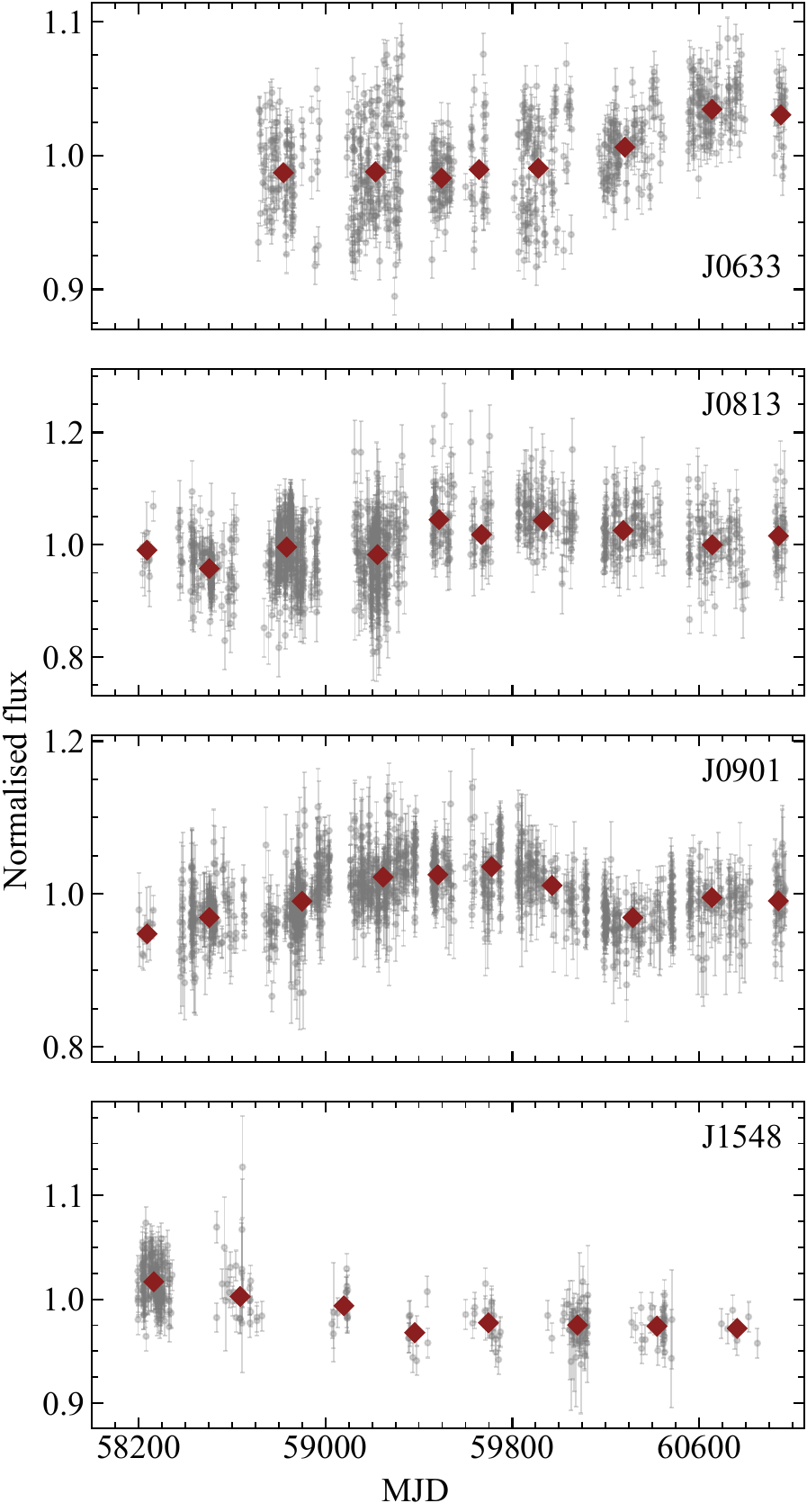}
\caption{ZTF multi-band light curves of four dC stars that exhibit clear long-term modulation, interpreted here as magnetic-activity cycles with variable starspot (or faculae) patterns. Each of these stars is also a short-period variable and thus a binary candidate. The panels are labeled with abbreviated J2000 identifiers that are unique within Table~\ref{tab:candidates}, and the red diamonds represent seasonal weighted means.  See Section~5.4 for a brief discussion.}
\label{fig:long_var}
\end{figure}

Figure~\ref{fig:long_var} displays the unprocessed light curves of four dC stars that are manifestly variable over seven years of available ZTF data. Despite the long-term coverage, the baselines are insufficient to quantify cyclical behavior in any of these stars, and thus neither detailed modeling nor period determination is attempted here.  Instead, the secular changes are quantified through a simple comparison with a constant-flux model.  A third-order polynomial is fitted using weighted least squares, after centering the time axis on the weighted-mean epoch to improve numerical conditioning. This polynomial is used purely as a generic description of smooth long-term behaviour, capturing gentle curvature or drift without imposing a specific light curve morphology.

To provide a quantitative summary of significance of the secular modulation, the $\upchi^{2}_\upnu$ values of the constant and cubic fits are compared.  This is not considered a formal test for long-term variability, as the candidates are selected visually where the multi-year trends are readily apparent in the light curves.  Nevertheless, in each of the four cases the polynomial fit improves the $\upchi^{2}_\upnu$ over a constant flux model by around 20--50\,per cent, and indicates these trends are real.  It is thus likely that other long-term, stellar activity trends exist in the data, but searching for these is beyond the scope of this work.



\begin{table}
\caption{Dwarf carbon star candidate variables with most likely periods $p$ and sine wave semi-amplitudes $A$, ordered by decreasing peak frequency significance $S \equiv -\log_{10}[X(z)]$.  The last column $R$ is the ratio of the adopted frequency power to the bootstrap threshold power corresponding to a 99.9\,per cent false-alarm probability.}
\label{tab:candidates}
\setlength{\tabcolsep}{4pt}
\begin{tabular}{lcllrr}


\hline

SDSS ID                         &$G$    &$p$            &$A$        &$S$    &$R$\\  
                                &(mag)  &(d)            &(\%)       &       &\\     

\hline


J234130.75+151943.4             &17.8   &0.13433692(5)  &5.87(8)    &$>99$  &27\\   
J081338.28+112321.2             &18.0   &2.52071(5)     &3.3(2)     &85     &9.0\\  
J125017.90+252427.6             &16.5   &2.9213(2)      &1.52(9)    &72     &7.4\\  
J081546.66+095952.7             &18.2   &1.18806(2)     &2.7(2)     &56     &6.1\\  
J151803.49+273601.9             &17.8   &1.04969(2)     &1.26(9)    &55     &6.4\\  
J063313.18+840900.1             &16.0   &0.310820(1)    &2.2(1)     &54     &6.1\\  
J151909.78+254218.1             &19.2   &7.699(1)       &2.1(2)     &32     &3.9\\  
J163718.64+274026.5             &17.5   &1.2280(9)      &0.7(1)     &31     &2.9\\  
J153059.25+451200.5             &16.7   &13.59(6)       &1.3(2)     &27     &3.6\\  
J090128.28+323833.5             &17.4   &2.1214(1)      &1.0(1)     &26     &3.6\\  
J154859.71+341821.7             &17.1   &3.33(1)        &1.2(2)     &15     &2.0\\  
J165902.30+250549.1             &19.4   &0.575360(6)    &1.7(1)     &14     &2.2\\  
J122357.67+550151.7$^{\rm a}$   &17.6   &0.30689(6)     &0.46(6)    &12     &1.2\\  
J094359.02+084323.7$^{\rm a}$   &18.4   &2.0960(1)      &1.8(3)     &8.8    &1.5\\  
J074908.53+284343.3             &20.3   &0.59430(1)     &3.4(5)     &7.9    &2.2\\  
J073805.72+274148.8             &18.4   &2.40(2)        &0.9(3)     &7.4    &1.4\\  
J080806.06+293634.1             &18.4   &2.60(1)        &0.8(2)     &6.9    &1.4\\  
J004706.76+000748.7$^{\rm a}$   &18.5   &12.574(4)      &1.3(2)     &6.5    &1.2\\  
J100132.62+020402.7             &19.4   &2.6883(3)      &2.2(4)     &6.5    &1.2\\  
J120246.02+541929.2             &19.1   &0.53514(1)     &1.6(2)     &6.3    &1.3\\  
J012150.42+011301.3             &17.1   &1.156(7)       &0.6(2)     &6.1    &1.2\\  
J112633.94+044137.6$^{\rm a}$   &16.8   &0.409532(6)    &1.1(9)     &6.1    &1.3\\  
J120031.66+035002.8             &20.2   &0.1688(3)      &9(3)       &6.1    &1.1\\  
J085953.59+092345.0             &19.5   &1.297(7)       &3.0(9)     &6.0    &1.1\\  
J112801.67+004034.6$^{\rm b}$   &18.9   &1.2240(6)      &3.9(7)     &5.9    &0.9\\  
J143100.09+131545.1$^{\rm a}$   &18.9   &1.1169(8)      &1.6(3)     &5.5    &1.1\\  
J074638.21+400403.4             &19.6   &0.3734(4)      &2.0(6)     &4.6    &1.2\\  
J081807.44+223427.5$^{\rm a}$   &15.6   &0.202323(2)    &0.26(4)    &4.0    &1.1\\  
J073341.27+222758.7$^{\rm a}$   &18.1   &0.6099(4)      &0.5(1)     &3.8    &1.0\\  
J083025.90+194700.9$^{\rm a}$   &20.0   &2.603(2)       &2.0(4)     &3.5    &1.1\\  
J083026.59+204023.3$^{\rm a}$   &19.6   &2.7111(3)      &1.9(3)     &3.5    &1.2\\  



\hline
\end{tabular}

{\em Notes.}
$^{\rm a}$At least one competing alias within 5\,per cent of peak amplitude.
$^{\rm b}$Eclipse.

\end{table}

\subsection{Discovery of an unexpected eclipse}

One of the key results of this work is the discovery of the first eclipsing dC star SDSS\,J112801.67+004034.6 (hereafter J1128), which offers a rare opportunity to constrain stellar parameters for an object in this class of likely old and metal-poor stars.  The nature of the eclipse is deduced from i) the periodogram of the unclipped light curve, which exhibits at least six harmonics of $f_1=0.8170$\,d$^{-1}$, most of which are above the adopted false-alarm threshold and with higher amplitude than $f_1$, and ii) light curves that are phase-folded on the periods corresponding to the higher periodogram peaks, all of which show distinct, time-dependent clumping below the otherwise-flat baselines in flux.  The subsequent analysis does not clip any of the ZTF data that pass other quality checks.

To characterize the eclipse, a box-fitting least squares analysis is performed on the $g$-, $r$-, and $i$-band light curves. The single eclipse is cleanly detected independently in each of the three light curves, with no significant evidence of a secondary eclipse.  Furthermore, across the full eclipse duration, there are no data points that return to the unocculted continuum in any of the three filters, indicating that this dimming feature is astrophysical and not an artefact of the phase-folding.  The derived ephemeris is:

\begin{flushleft}
HMJD $=58217.82517(3) + 1.223975(5)E$
\end{flushleft}

\medskip
\noindent
with errors derived by Monte Carlo methods assuming the rectangular geometry of box least squares.

\begin{figure}
\centering
\includegraphics[width=\columnwidth]{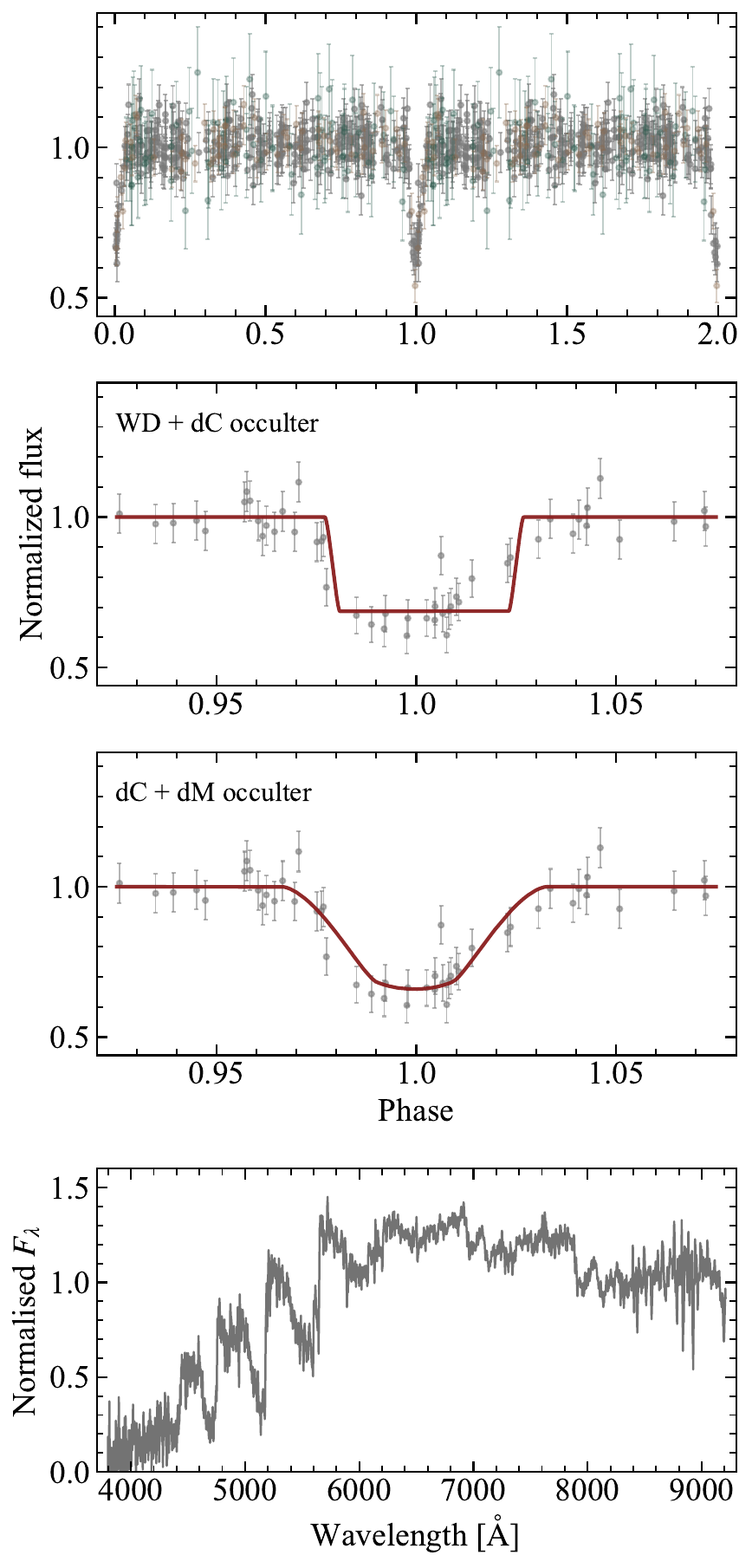}
\caption{The top panel shows the full $r$-band light curve of J1128 as seen by ZTF, plotted in grey and folded on the most likely period of 1.2240\,d.  Although the $g$- and $i$-band data are too sparse to produce unique light curves, these independent data corroborate the timing and the depth of the dimming, and are overplotted in dark green and brown, respectively.  The two middle panels each plot a simulated eclipse (see text): a white dwarf that is occulted by a dC star (upper), and a dC star that is occulted by an M dwarf (lower).  The bottom panel shows the SDSS spectrum, which is consistent with a single star and thus an SB1 type binary.  \phantom{THIS IS A BUNCH OF PHANTOM TEXT TO GET THE FIGURE TO STICK.  THIS IS A BUNCH OF PHANTOM TEXT TO GET THE FIGURE TO STICK.  THIS IS A BUNCH OF PHANTOM TEXT TO GET THE FIGURE TO STICK.  THIS IS A BUNCH OF PHANTOM TEXT TO GET THE FIGURE TO STICK.  THIS IS A BUNCH OF PHANTOM TEXT TO GET THE FIGURE TO STICK.  THIS IS A BUNCH OF PHANTOM TEXT TO GET THE FIGURE TO STICK.}}
\label{fig:eclipse}
\end{figure}

Further analysis is carried out on the $r$-band light curve, which is the only one that provides a well-constrained eclipse depth; observations in the other two filters lack sufficient phase coverage across the dimming event.  To constrain the unocculted flux continuum, the eclipse region is masked and a single-pass $1\upsigma$ median clip is applied to the remaining data points.  The light curve is then re-normalized by the resulting median, and subsequently phase-folded on the best period.  The result is shown in Figure~\ref{fig:eclipse}, where a striking feature is the eclipse depth, which is at least two orders of magnitude too deep to be caused by a white dwarf occulter, where typically $(R_\star/R_\odot)^2\approx2\times10^{-4}$, thus ruling out the secondary as an extrinsic donor of carbon.

Using {\sc lcurve} \citep{Copperwheat2010}, two distinct light curve eclipses are simulated and shown in Figure~\ref{fig:eclipse}, both of which assume $i=90\degr$ and linear limb darkening in all components.  The first model is a white dwarf that is eclipsed by the dC star, where a good fit is achieved with $R_1=0.015$\,R$_\odot$, $R_2=0.5$\,R$_\odot$, and a brightness ratio such that the white dwarf contributes around 70\,per cent in the ZTF $r$ band.  This possibility can be immediately ruled out, as the bottom panel of Figure~\ref{fig:eclipse} demonstrates there is no evidence of a white dwarf in the SDSS spectrum at any wavelength.  The second model utilizes the same dC star radius (in this case as $R_1$), but using an occulting M dwarf with $R_2=0.3$\,R$_\odot$, and which contributes less than 5\,per cent of the flux in the $r$ band.  

All the data are consistent with this second simulated light curve.  Beyond the eclipse morphology itself, both the optical spectrum and the lack of a secondary eclipse require that the dC star is eclipsed by a dark companion.  The best fit simulated parameters are given in Table~\ref{tab:eclipse}, and leave clear room for improvement using future observations.  For example, a full eclipse with $R_2/R_1=0.54$ should yield $\Delta F=0.29$ as compared to 0.32 in the simulation, and thus the eclipse may be grazing.  Given the relatively noisy photometry and the sparse coverage of the eclipse itself, the results are likely to be correct in a broad sense, where improved accuracy requires better data.

\begin{table}
\centering
\caption{Derived parameters from the $r$-band eclipse light curve of J1128.  Errors are not given as the fitted model assumes $i=90$\degr.}
\label{tab:eclipse}
\begin{tabular}{lr}

\hline

Quantity & Value\\

\hline

$p$ (d)             & 1.2240\\
$t_4-t_1$ (h)       & 1.94\\
$t_3-t_2$ (h)       & 0.57\\
$R_1/a$             & 0.134\\
$R_2/a$             & 0.073\\
$R_2/R_1$           & 0.54\\
$\Delta F$          & 0.32\\

\hline

\end{tabular}
\end{table}

\section{DISCUSSION}

The overall results of this study are compared to previous work on the photometric variability and associated duplicity of dC stars.  Based on this outcome, the orbital period distribution and constraints on the fraction of short-period dC binaries are detailed, with some caveats on completeness and accuracy, and possibly implications for mass-transfer efficiency as a function of orbital period.  The evidence for tidal-locking, rejuvenated spins, and magnetic activity is reviewed and substantiated. Lastly, the implications of the eclipsing binary J1128 are discussed, with an emphasis on the need for additional data.

\subsection{An assessment of the \texorpdfstring{$gri$}{{\it gri}} combined light curve approach}

In the subsequent section is a comparison of the methodology adopted here with a previous dC star variability study using ZTF, where analysis was carried out using individual filter light curves \citepalias{Roulston2021}.  Thus, the band-combined light curve analysis done here merits further scrutiny to provide a more informed discussion of the two sets of results.

The primary concern of combining (re-normalized) photometric data in different bands is the potential dilution of a real signal in one band that is absent in one or more other bands.  To assess this possibility, all the individual filter light curves for Table~\ref{tab:candidates} sources are analyzed using the Sample I methodology, with informative results.  Among those dC stars with the highest amplitude variations, there are five objects where the peak signal is recovered in all three filters, and two more where the signal is recovered both in $r$ and either $g$ or $i$.  For these seven dC stars, the means and standard deviations of the amplitude ratios are such that: i) $A_r/A_g=0.84\pm0.10$, where the $g$-band amplitude is always larger, and ii) $A_i/A_r=0.90\pm0.10$, where the $i$-band amplitudes are usually weaker but can be modestly stronger than the $r$ band.

Because the bulk of ZTF data are taken using the $r$-band filter, and assuming the above correlations hold for the wider dC population, it can be inferred that adding the $g$ band should only amplify the signal sought, and adding the $i$ band is unlikely to dilute any signal, but can in fact strengthen it.  In corroboration of these inferences, in Table~\ref{tab:candidates} there are six sources for which all three individual light curves have no periodogram peaks that are significant, and yet the $gri$-combined analysis passes all the requirements.  Overall, these outcomes and considerations are taken as evidence that the band-combined approach is likely to be more sensitive than the analysis of individual bandpass light curves, at least for this particular type of astrophysical signal, which is either (dark) starspots or (bright) faculae on late-type main-sequence stars.

\subsection{Direct comparison with \texorpdfstring{\citetalias{Roulston2021}}{R21}}

Sample II consists of 34 candidate dC stars with photometric periods previously reported on the basis of their ZTF light curves \citepalias{Roulston2021}.  Those authors use DR5 with only CATFLAGS $=0$ data filtering, and then identify variability using Lomb--Scargle periodogram analysis of light curves constructed from single bandpass filters.  Here, these 34 candidates are independently assessed using the Sample I methodology: in particular the $gri$ band-combined approach, with additional data filtering, longer baselines of DR24, and checking each dC star spectrum by eye to confirm its nature.

In addition to the previous discussion, there are two methodological differences that are noteworthy.  First, in \citetalias{Roulston2021} there is no explicit {\em Gaia}-based neighbour check, which can result in source confusion and time-variable flux contamination within the ZTF aperture or light curve association, and can bias period recovery \citep{Guidry2021}.  Second, they adopt a relatively narrow alias association of $\pm0.005$\,d$^{-1}$ for each candidate periodogram peak (cf.\ here all peaks within $\pm0.02$\,d$^{-1}$ are considered possible aliases), which risks classifying the effects of spectral leakage as genuine variability \citep{VanderPlas2018}.  These choices motivate a more careful analysis with stricter crowding control and window function vetting.

Applying the Sample I methodology to Sample II yields results summarized in Table~\ref{tab:r21}. Of the 34 reported variables in \citetalias{Roulston2021}, only 12 (35\,per cent) have verified periods that agree to within a reasonable degree with those found here.  In the comparison, the peak frequency periodogram signals and the corresponding periods have been compared in a relatively generous fashion, because starspot groups can grow and disappear on the timescales relevant to the differences between ZTF DR5 and DR24.  Nevertheless, despite adopting a relatively loose condition of $\pm10$\,per cent for agreement, the remaining 22 candidates have modestly to drastically different interpretations for several reasons noted in the table, and detailed further below.  

For seven objects, the signals adopted in \citetalias{Roulston2021} do not correspond to the strongest frequency peaks in the band-combined periodograms.  Six of these are consistent with aliases of varying rank in strength, most of which are below the second strongest, but in principle are associated with the strongest peak found in the band-combined light curve and thus remain candidate variables.  The seventh object has a reported period that is an alias of the observing cadence and can be firmly ruled out.  Thus, 19 of the 34 sources reported as variable are likely to be genuine variables and likely short-period binaries.  For the remaining 15 sources, all fail one or more of the criteria used for Sample I.

Ten sources show no clear frequency peaks above the $z_\star$ and bootstrap false-alarm thresholds based on DR24 data, or have peak frequencies clearly associated with the nightly observing cadence (and interaction with the lunar cycle).  Furthermore, analysis of their individual bandpass light curves does not change this result: no source has any unmasked peak above the 99.9\,per cent false alarm in any filter, not even when a DR5-like cut is made to the data in the light curves.  Three dC stars have light curves that are contaminated by sources that fail to satisfy the first criterion listed in Section~2.2:  LAMOST\,J091458.10+215639.6 and SDSS\,J152504.52+322511.3 both have common-proper motion companions within 3\,arcsec and which have $\Delta m<1.0$\,mag; LAMOST\,J192355.97+445833.2 sits within 5\,arcsec of a background star with $\Delta m<2.0$\, mag.  Lastly, two objects are not confirmed dC stars: LAMOST\,J041605.11+502828.5 is a main-sequence star based on {\sc obsid} 696104148; and LAMOST\,J121007.00+584318.5 is not found in any of the LAMOST public data releases (through DR11), and is conspicuously absent from DR4 where the source is reported by \citet{Li2018}; also, there are two {\em Gaia} sources within 2\,arcsec of its position (with $G=17.2$ and 18.3\,mag, respectively, both faint for this survey).

\begin{table}
\caption{Results of an independent analysis of 34 dC star candidates published as photometrically variable in ZTF DR5 \citepalias{Roulston2021}. The fourth column contains two characters that are each either `+' or `$-$' based on the Sample I methodology: the first indicates whether the \citetalias{Roulston2021} period is identified, and the second indicates whether the star passes the variability criteria.}

\label{tab:r21}
\begin{tabular}{lcc@{}cl}

\hline

Source ID
&\multicolumn{2}{c}{Period} 
&
&Notes\\

\cline{2-3}
\rule{0pt}{2.5ex} 

&R21 
&This study 
&
&and Refs.\\

&(d) 
&(d)
&
&\\

\hline

Variable, period confirmed (12):\\   
SDSS\,J004706.76+000748.7       &12.61      &12.57      &++     &1,a\\      
LAMOST\,J050240.82+402323.6     &4.428      &4.433      &++     &2\\        
LAMOST\,J074447.67+513832.0     &1.535      &1.535      &++     &2\\        
SDSS\,J081157.14+143533.0       &0.750      &0.750      &++     &1\\        
LAMOST\,J123045.53+410943.8     &0.883      &0.882      &++     &3\\        
LAMOST\,J130359.18+050938.6     &1.841      &1.845      &++     &2\\        
SBSS\,1310+561                  &5.188      &5.249      &++     &4,b\\      
SBSS\,1517+502                  &0.302      &0.302      &++     &6,b\\      
SDSS\,J153059.25+451200.5       &13.59      &13.59      &++     &1,a\\      
SDSS\,J163718.64+274026.5       &1.228      &1.228      &++     &1,a\\      
LAMOST\,J220809.97+251729.9     &0.423      &0.423      &++     &5\\        
SDSS\,J234130.75+151943.4       &0.134      &0.134      &++     &1,a\\      

\\
Variable, period distinct (7):\\    
LAMOST\,J023530.65+022518.6     &1.660      &2.488      &$-$+   &2,c\\      
LAMOST\,J025414.25+262154.0     &0.190      &0.470      &$-$+   &5,c\\      
LAMOST\,J062558.33+023019.4     &7.608      &0.884      &$-$+   &5,c\\      
SDSS\,J120246.02+541929.2       &1.155      &0.535      &$-$+   &1,a,c\\    
SDSS\,J122357.67+550151.7       &0.336      &0.307      &$-$+   &1,a,d\\    
LAMOST\,J140953.08-061141.8     &0.320      &8.494      &$-$+   &2,c\\      
SDSS\,J165902.30+250549.1       &0.288      &0.575      &$-$+   &1,a,c\\    

\\
No clear periodicity (10):\\           
LAMOST\,J013119.05+372025.2     &1.377      &           &$-$$-$ &2,c,e\\    
HE\,0930$-$0018                 &1.157      &           &$-$$-$ &7,e\\      
SDSS\,J094026.29+362548.7       &1.957      &           &$-$$-$ &1,d,e\\    
SDSS\,J120853.36$-$000847.3     &0.351      &           &$-$$-$ &1,e\\      
SDSS\,J133123.61+482624.4       &0.204      &           &$-$$-$ &1,e\\      
SDSS\,J141515.25+514128.2       &0.273      &           &$-$$-$ &1,e\\      
SDSS\,J151144.60+385910.7       &0.336      &           &$-$$-$ &1,d,e\\    
SDSS\,J151542.92+520145.6       &0.333      &           &$-$$-$ &1,d,e\\    
SDSS\,J152434.12+444956.1       &0.252      &           &$-$$-$ &8,d,e\\    
SDSS\,J153532.93+011016.4       &0.174      &           &$-$$-$ &1,e\\      

\\
Contaminated photometry (3):\\ 
LAMOST\,J091458.10+215639.6     &1.236      &           &$-$$-$ &2,f\\    
SDSS\,J152504.52+322511.3       &0.137      &           &$-$$-$ &1,f\\    
LAMOST\,J192355.97+445833.2     &0.146      &           &$-$$-$ &2,f\\    

\\
Not dC stars (2):\\
LAMOST\,J041605.11+502828.5     &6.808      &           &$-$$-$ &2,g\\
LAMOST\,J121007.00+584318.5     &0.184      &           &$-$$-$ &2,f,g\\
                                                                                                        
\hline

\end{tabular}

{\em Notes and references.} (1) \citet{Green2013}; (2) \citet{Li2018}; (3) \citet{Bai2016}; (4) \citet{Rossi2011}; (5) \citet{Ji2016}; (6) \citet{Liebert1994}; (7) \citet{Christlieb2001}; (8) \citet{Downes2004}; (a) Table~\ref{tab:candidates}; (b) duplicity previously established; (c) \citetalias{Roulston2021} alias of strongest peak; (d) \citetalias{Roulston2021} alias of sidereal day as verified by window function or ZTF source classification work (Section~3.2); (e) no periodogram peaks above the 99.9\,per cent false-alarm threshold power; (f) fails first criterion in Section~2.2; (g) not confirmed dC star (see Section~5.2).

\end{table}

\subsection{An evolutionary framework for variable dC stars}

The detected photometric variability is interpreted here using the framework of tidal locking in post-common envelope binaries.  It is most probable that any relatively close companion to a dC star is an unseen white dwarf whose progenitor provides both the envelope to cause orbital contraction via dynamical friction, as well as an external source of carbon for mass transfer \citep{Dearborn1986,Heber1993,Liebert1994}. The latter requires that the common envelope is formed after the third dredge-up, and thus relatively late in the evolution of an AGB star.  

In this picture it might be expected that post-common envelope dC stars have an orbital period distribution with a higher proportion of longer periods as compared to the orbits of similar binaries with carbon-normal, main-sequence stars \citep{Kool1995}.  White dwarfs with closely orbiting M dwarfs have period distributions that are either log-normal and dominated by systems with $p<1$\,d \citep{Nebot2011}, or log-uniform over the range $0.1<p<2$\,d \citep{Ashley2019,Shariat2026}.  In either case there appears to be a dearth of systems with periods longer than 1--2\,d in these binaries with unpolluted and unevolved companions.  

\begin{figure}
\centering
\includegraphics[width=\columnwidth]{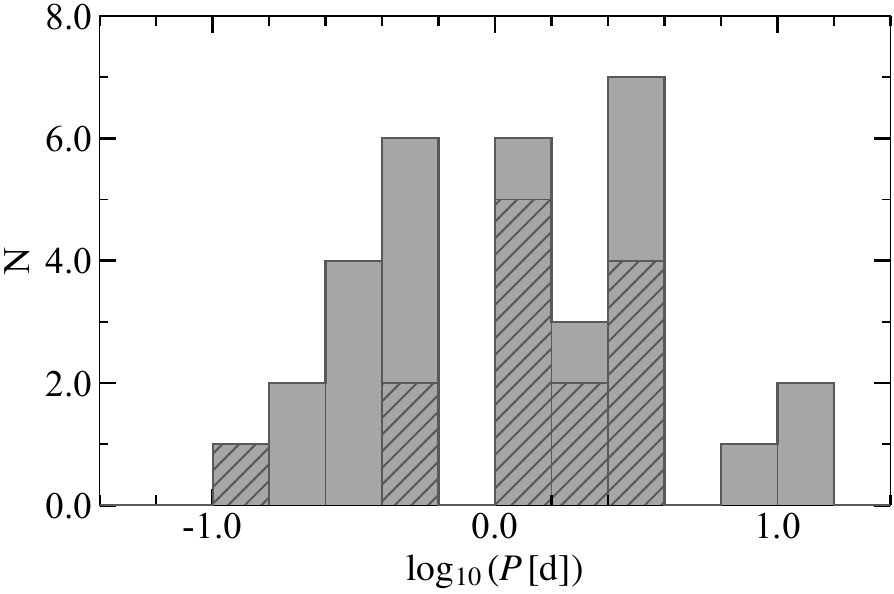}
\caption{Distribution of Table~\ref{tab:candidates} periods.  The hatching represent those stars whose SDSS spectrum shows an unambiguous emission line at H$\upalpha$.  Based on a superficial comparison to white dwarf plus main-sequence binaries that are carbon-normal, there may be a dearth of $p<1$\,d periods and a modest excess of $p\ga1$\,d periods for dC stars.  However, the sample is biased and incomplete, as most stars have no radial velocity data so caution is warranted.}
\label{fig:phist}
\end{figure}

In Figure~\ref{fig:phist} plots the 31 relatively high-confidence dC star photometric periods that are likely linked to their true orbital periods (Table~\ref{tab:candidates}).  While differential rotation may cause a group of spots to rotate somewhat more slowly than the orbit duration in tidally-locked binaries, they should not differ by more than around 25\,per cent based on spots and differential rotation in the Sun and cooler stars \citep{Barnes2005}, as well as a handful of dC stars with independently determined photometric and radial velocity periods \citep{Whitehouse2021}.  Such modest differences are insufficient to alter the character of the inferred dC period distribution in a significant manner, even in a worst-case scenario.

The distribution shown in Figure~\ref{fig:phist} should be considered biased and incomplete.  The vast bulk of variable sources have no radial velocity confirmation of duplicity, the sample is relatively faint, and as previously discussed, ground-based photometric monitoring makes it challenging to unambiguously identify variability on timescales similar to nightly cadences.  But the largest source of uncertainty may be the unknown fraction of dC stars that become sufficiently magnetically active in post-common envelope binaries, and any dependence on orbital period, age, or metallicity.  There is at least one dC star with a short-period orbit detected via radial velocity \citep[LSR\,J2105+2514, $p=0.15$\,d;][]{Whitehouse2023}, but no photometric variation in public surveys, nor any emission lines in its spectrum. Table~\ref{tab:candidates} stars also show a wide range of activity strengths, including no emission lines in their SDSS spectra; this range may simply extend to undetectable photometric variation.

Nevertheless, superficially, the distribution appears to have relatively fewer objects at orbital periods in the 2--10\,h range, which tend to dominate the (bias-uncorrected) distributions measured for white dwarf plus M dwarf systems.  Also, it may be the case that the dC star distribution has a modestly higher proportion of periods beyond 2\,d, which is a relatively barren region for M dwarf companions \citep{Nebot2011,Ashley2019,Shariat2026}.

The presence of chromospheric emission lines is an independent line of evidence that helps strengthen the link between the photometric period distribution and the orbital period distribution.  Among the 31 candidate variable dC stars, there are 14 objects whose SDSS spectra exhibit clear evidence for emission at H$\upalpha$, and which are highlighted by the hatched columns in Figure~\ref{fig:phist}.  The correlation is suggestive, and while nothing conclusive can be inferred, there appears to be a decent prevalence of systems that require a source of rejuvenated spin, and are thus likely to be in tidally-locked binaries.  If one were only to interpret the distribution of these 14 periods, then similar inferences might be made in comparison to short-period M dwarf companions to white dwarfs, i.e.\ the dC stars appear to have a modestly higher proportion of longer periods, and thus this preliminary orbital period distribution may be indicative of their actual properties.

In this study there are 31 relatively strong variability candidates among 879 confirmed dC stars with sufficient ZTF data to analyze.  This yields a lower limit fraction $f\geq0.035$ (31/879), and is only a crude constraint on the post-common envelope binary fraction of the dC star population at large.  This value can be compared to that estimated by \citet{Roulston2021}, who report 0.036 (34/944), but is independently re-evaluated as 0.020 (19/942) here, assuming all the Table~\ref{tab:r21} sources for which variability seems reasonable.  Hence it appears the band-combined strategy is roughly 75\,per cent more efficient in the confident identification of variable dC stars.

\subsection{Spin and dynamo rejuvenation from tidal locking}

Because the dC star population is dominated by relatively old stars orbiting within the Galactic halo and thick disk \citep{Farihi2025}, they are expected to be slow rotators in isolation.  Even in the case that their spins are increased by mass transfer from a carbon-rich companion as has been previously suggested \citep{Green2019}, they will efficiently spin down again within 1\,Gyr without further influence, as do all solar-type stars above the fully convective boundary \citep[e.g.][]{Kraft1967,Barnes2003}. In contrast, tidal locking is an enduring physical mechanism to increase and sustain rapid rotation, and for periods of hours to days is sufficient to rejuvenate a dynamo and associated stellar activity \citep[e.g.][]{Vidotto2014}. 

Other possibilities to account for the properties of photometrically variable dC stars can be decisively ruled out, based on multiple lines of evidence \citep[see figs.~4 and 6 in][]{Whitehouse2021}.  First, their light curves all have a single minimum, with no correlation between the period and amplitude of variability, and thus are inconsistent with ellipsoidal modulation.  Second, while there are a small number of white dwarf companions that have been detected in the optical or the ultraviolet, most companions are dark and thus irradiation of the stellar surface cannot be responsible for the observed photometric modulation.  Third, in those systems with short orbital periods verified by radial velocity monitoring, the orbital period is typically shorter than the photometric period, consistent with differential rotation and starspots at modest latitudes (see Section~5.3).  Furthermore, in at least one source, the photometric variability has been shown to vary in amplitude and modestly drift in phase and frequency over several months, consistent with changing starspot groups.  Fourth, there is an observed correlation between strong chromospheric emission lines in the dC stars and photometric variability, consistent with a rejuvenated dynamo. 

Further support for this picture is given in Figure~\ref{fig:long_var}, where all four sources with obvious, long-term, light-curve trends are also among the most confident, short-period variable candidates in Table~\ref{tab:candidates}.  The trends in flux over many years are analogous to the solar magnetic activity cycle, which peaks roughly every 11\,yr and produces high sunspot counts.  Of the six dC stars with published X-ray detections \citep{Green2019}, five (J0901, J1015, J1250, J1548, and J1637) are in Table~\ref{tab:candidates}, and two exhibit clear long-term photometric variability in Figure~\ref{fig:long_var}.  Of these five, all but J1548 have been previously identified as short-period binaries, while the sixth X-ray source (CBS\,311) is not in Sample I but is also a known short-period binary \citep{Margon2018,Whitehouse2021,Roulston2021}.  The preponderance of evidence supports the interpretation of dC star photometric variability as starspots that result from rejuvenated spins and magnetic activity in tidally-locked binaries.  

With additional monitoring data in the coming years via ZTF and the Rubin Observatory, it may be possible to directly constrain the periodic (magnetic) activity cycles of rapidly-rotating dC stars.  These could be compared to what is known for normal G and K-type dwarfs, which occupy the same part of the HR diagram as the dC stars \citep{Whitehouse2023}.  Recent studies have called into question the possibility of distinct evolutionary tracks for younger, active, and rapidly rotating solar-type stars as compared to those that are older, less active or inactive, and more slowly rotating \citep{Bohm2007,Boro2018,Chahal2025}.  Placing the dC stars into this context could shed light on basic stellar structure and dynamo evolution in a set of likely metal-poor stars, likely empirical constraints that are not represented in the solar neighborhood G and K-type stars.

In one sense, the rapidly-rotating dC stars likely provide a glimpse of the early spin and dynamo evolution in metal-poor stars, which would be otherwise impossible to observe today.  Notably, only nine of the 12 most significant variable candidates in Table~\ref{tab:candidates} exhibit clear emission at H$\upalpha$.  Furthermore, J2341 has the strongest rotation signal and has the shortest known period among dC stars at just 7.4\,h \citep{Whitehouse2021,Roulston2021}, but has relatively weak emission at H$\upalpha$.  Lastly, J0633 has a relatively high significance for its periodogram peak, has clear long-term variability in Figure~\ref{fig:long_var}, but absolutely no evidence for emission lines in its spectrum.  Hence there remains more to learn about the spin and dynamo evolution of rejuvenated dC stars.

\subsection{Possibilities for the unexpected eclipse of J1128}

The primary star in this system is an unambiguous halo star.  It has the largest radial velocity among 1220 candidate dC stars with SDSS spectroscopy \citep{Green2013}, and has been previously interpreted as a possible runaway and candidate supernova ejecta \citep{Plant2016}.  However, it is bound to the Galaxy and has a relatively high-velocity halo orbit with $(U,V,W) = (16,-435,356)$\,km\,s$^{-1}$, but is otherwise not unusual as it sits among nearly 200 dC stars with total space velocities exceeding 400\,km\,s$^{-1}$ \citep{Farihi2025}.

The eclipse profile of J1128 implies that the companion is a low-mass star or even perhaps a brown dwarf, but by no means a white dwarf.  It has RUWE $=0.998$ in {\em Gaia} DR3 and thus currently shows no astrometric indication of a tertiary star, and no ultraviolet detection at its position in {\em GALEX}.  In principle, an isolated binary with no possible extrinsic donor of carbon would point to a primordial origin for the carbon-enhanced atmosphere of one or both stars in the eclipsing binary.  

However, studies of solar-type stars have shown that the overwhelming majority of close binaries with $p_<3$\,days have tertiary companions \citep{Tokovinin2006,Laos2020}.  Theoretical work suggests it is challenging to form two stars at these separations, and it is likely such observed systems were born further apart and a tertiary star has hardened the inner binary by dynamical mechanisms such as Kozai-Lidov, which can then widen the orbit of the tertiary \citep{Moe2018}.  It is therefore entirely possible that J1128 is currently part of a triple system, which may come into evidence with improved {\em Gaia} astrometry in subsequent data releases, or by radial velocity monitoring of the inner pair.  A distinct possibility is that such a tertiary can be ejected, especially if the tertiary loses substantial mass as expected for an AGB star which donated carbon to the inner binary \citep{Shappee2013}.  It thus remains possible that the carbon enhancement in this system is via mass transfer, but would have to eschew a common envelope with three stars, which would be unstable.

\section{CONCLUSION}

Using time-domain ZTF photometry of 1063 confirmed dC stars, 31 candidate periodic variables are identified, most of which are new discoveries.  The bulk of these have $p<3$\,d, with three objects exhibiting longer periods up to 13\,d.  These signals are naturally explained by tidally-locked rotation in short-period, post-common envelope systems, and corroborated by a subset of 14 stars that exhibit chromospheric emission consistent with rejuvenated spins and internal dynamos, and a separate subset that has multi-year trends in the light curves, consistent with stellar activity cycles.  Under the assumption that all 31 are indeed binaries as described above, the orbital period distribution appears moderately shifted to longer periods as compared to carbon-normal, low-mass stellar companions to white dwarfs in post-common envelope binaries.  However, as the bulk of candidates lack radial velocity data, this preliminary result should be viewed with caution owing to bias, sensitivity, and incompleteness.  

The results here have been compared to prior work using ZTF and in many cases those earlier results are not confirmed.  It is a cautionary tale that ground-based photometric surveys contain many challenges in identifying weak signals toward faint objects such as the dC stars.  Lastly, the first eclipsing dC star J1128 is a puzzling discovery, as the light curve firmly rules out a white dwarf secondary and thus a prior carbon donor.  However, such a companion may still await detection and better data are paramount.  The detection of a secondary eclipse will directly yield an accurate radius ratio of the two stars, and combined the primary eclipse yield both stellar radii and the orbital inclination.  Spectroscopy is necessary to search for a second set of spectral lines in this system, and if detected can yield both the masses and radii independent of models.

\section*{Acknowledgements}

The authors thank an anonymous reviewer for feedback that helped improve the manuscript.  The team is grateful to J.~J.~Hermes for advice and discussions on the vagaries of ground-based, time-series photometry and analysis, J.~Guidry for guidance on best practices using ZTF data, and R.~R.~Rafikov for the basic dynamics of mass transfer from a tertiary source onto a compact binary.  This paper is based on observations obtained with the Samuel Oschin 48-inch and the 60-inch Telescopes at the Palomar Observatory as part of the Zwicky Transient Facility project, supported by the National Science Foundation under Grants No. AST-1440341 and AST-2034437.

\section*{Data availability}
All Zwicky Transient Facility data are available to download from the NASA / IPAC infrared science archive.


\bibliographystyle{mnras}
\bibliography{refs} 

\begin{thebibliography}{}
\makeatletter
\relax
\def\mn@urlcharsother{\let\do\@makeother \do\$\do\&\do\#\do\^\do\_\do\%\do\~}
\def\mn@doi{\begingroup\mn@urlcharsother \@ifnextchar [ {\mn@doi@}
  {\mn@doi@[]}}
\def\mn@doi@[#1]#2{\def\@tempa{#1}\ifx\@tempa\@empty \href
  {http://dx.doi.org/#2} {doi:#2}\else \href {http://dx.doi.org/#2} {#1}\fi
  \endgroup}
\def\mn@eprint#1#2{\mn@eprint@#1:#2::\@nil}
\def\mn@eprint@arXiv#1{\href {http://arxiv.org/abs/#1} {{\tt arXiv:#1}}}
\def\mn@eprint@dblp#1{\href {http://dblp.uni-trier.de/rec/bibtex/#1.xml}
  {dblp:#1}}
\def\mn@eprint@#1:#2:#3:#4\@nil{\def\@tempa {#1}\def\@tempb {#2}\def\@tempc
  {#3}\ifx \@tempc \@empty \let \@tempc \@tempb \let \@tempb \@tempa \fi \ifx
  \@tempb \@empty \def\@tempb {arXiv}\fi \@ifundefined
  {mn@eprint@\@tempb}{\@tempb:\@tempc}{\expandafter \expandafter \csname
  mn@eprint@\@tempb\endcsname \expandafter{\@tempc}}}

\bibitem[\protect\citeauthoryear{{Arentsen}, {Starkenburg}, {Shetrone}, {Venn},
  {Depagne}  \& {McConnachie}}{{Arentsen} et~al.}{2019}]{Arentsen2019}
{Arentsen} A.,  {Starkenburg} E.,  {Shetrone} M.~D.,  {Venn} K.~A.,  {Depagne}
  {\'E}.,   {McConnachie} A.~W.,  2019, \mn@doi [\aap]
  {10.1051/0004-6361/201834146}, \href
  {https://ui.adsabs.harvard.edu/abs/2019A&A...621A.108A} {621, A108}

\bibitem[\protect\citeauthoryear{{Ashley}, {Farihi}, {Marsh}, {Wilson}  \&
  {G{\"a}nsicke}}{{Ashley} et~al.}{2019}]{Ashley2019}
{Ashley} R.~P.,  {Farihi} J.,  {Marsh} T.~R.,  {Wilson} D.~J.,   {G{\"a}nsicke}
  B.~T.,  2019, \mn@doi [\mnras] {10.1093/mnras/stz298}, \href
  {https://ui.adsabs.harvard.edu/abs/2019MNRAS.484.5362A} {484, 5362}

\bibitem[\protect\citeauthoryear{{Bai} et~al.,}{{Bai} et~al.}{2016}]{Bai2016}
{Bai} Y.,  et~al., 2016, \mn@doi [Research in Astronomy and Astrophysics]
  {10.1088/1674-4527/16/7/107}, \href
  {https://ui.adsabs.harvard.edu/abs/2016RAA....16..107B} {16, 107}

\bibitem[\protect\citeauthoryear{{Baluev}}{{Baluev}}{2008}]{Baluev2008}
{Baluev} R.~V.,  2008, \mn@doi [\mnras] {10.1111/j.1365-2966.2008.12689.x},
  \href {https://ui.adsabs.harvard.edu/abs/2008MNRAS.385.1279B} {385, 1279}

\bibitem[\protect\citeauthoryear{{Barnes}}{{Barnes}}{2003}]{Barnes2003}
{Barnes} S.~A.,  2003, \mn@doi [\apj] {10.1086/367639}, \href
  {https://ui.adsabs.harvard.edu/abs/2003ApJ...586..464B} {586, 464}

\bibitem[\protect\citeauthoryear{{Barnes}, {Collier Cameron}, {Donati},
  {James}, {Marsden}  \& {Petit}}{{Barnes} et~al.}{2005}]{Barnes2005}
{Barnes} J.~R.,  {Collier Cameron} A.,  {Donati} J.-F.,  {James} D.~J.,
  {Marsden} S.~C.,   {Petit} P.,  2005, \mn@doi [\mnras]
  {10.1111/j.1745-3933.2005.08587.x}, \href
  {https://ui.adsabs.harvard.edu/abs/2005MNRAS.357L...1B} {357, L1}

\bibitem[\protect\citeauthoryear{{Bellm} et~al.,}{{Bellm}
  et~al.}{2019}]{Bellm2019}
{Bellm} E.~C.,  et~al., 2019, \mn@doi [\pasp] {10.1088/1538-3873/aaecbe}, \href
  {https://ui.adsabs.harvard.edu/abs/2019PASP..131a8002B} {131, 018002}

\bibitem[\protect\citeauthoryear{{B{\"o}hm-Vitense}}{{B{\"o}hm-Vitense}}{2007}]{Bohm2007}
{B{\"o}hm-Vitense} E.,  2007, \mn@doi [\apj] {10.1086/510482}, \href
  {https://ui.adsabs.harvard.edu/abs/2007ApJ...657..486B} {657, 486}

\bibitem[\protect\citeauthoryear{{Boro Saikia} et~al.,}{{Boro Saikia}
  et~al.}{2018}]{Boro2018}
{Boro Saikia} S.,  et~al., 2018, \mn@doi [\aap] {10.1051/0004-6361/201629518},
  \href {https://ui.adsabs.harvard.edu/abs/2018A&A...616A.108B} {616, A108}

\bibitem[\protect\citeauthoryear{{Chahal}, {Kamath}, {de Grijs}, {Montet}  \&
  {Chen}}{{Chahal} et~al.}{2025}]{Chahal2025}
{Chahal} D.,  {Kamath} D.,  {de Grijs} R.,  {Montet} B.~T.,   {Chen} X.,  2025,
  \mn@doi [\mnras] {10.1093/mnras/staf754}, \href
  {https://ui.adsabs.harvard.edu/abs/2025MNRAS.540..668C} {540, 668}

\bibitem[\protect\citeauthoryear{{Christlieb}, {Green}, {Wisotzki}  \&
  {Reimers}}{{Christlieb} et~al.}{2001}]{Christlieb2001}
{Christlieb} N.,  {Green} P.~J.,  {Wisotzki} L.,   {Reimers} D.,  2001, \mn@doi
  [\aap] {10.1051/0004-6361:20010814}, \href
  {https://ui.adsabs.harvard.edu/abs/2001A&A...375..366C} {375, 366}

\bibitem[\protect\citeauthoryear{{Copperwheat}, {Marsh}, {Dhillon},
  {Littlefair}, {Hickman}, {G{\"a}nsicke}  \& {Southworth}}{{Copperwheat}
  et~al.}{2010}]{Copperwheat2010}
{Copperwheat} C.~M.,  {Marsh} T.~R.,  {Dhillon} V.~S.,  {Littlefair} S.~P.,
  {Hickman} R.,  {G{\"a}nsicke} B.~T.,   {Southworth} J.,  2010, \mn@doi
  [\mnras] {10.1111/j.1365-2966.2009.16010.x}, \href
  {https://ui.adsabs.harvard.edu/abs/2010MNRAS.402.1824C} {402, 1824}

\bibitem[\protect\citeauthoryear{{Coughlin} et~al.,}{{Coughlin}
  et~al.}{2021}]{Coughlin2021}
{Coughlin} M.~W.,  et~al., 2021, \mn@doi [\mnras] {10.1093/mnras/stab1502},
  \href {https://ui.adsabs.harvard.edu/abs/2021MNRAS.505.2954C} {505, 2954}

\bibitem[\protect\citeauthoryear{{Dahn}, {Liebert}, {Kron}, {Spinrad}  \&
  {Hintzen}}{{Dahn} et~al.}{1977}]{Dahn1977}
{Dahn} C.~C.,  {Liebert} J.,  {Kron} R.~G.,  {Spinrad} H.,   {Hintzen} P.~M.,
  1977, \mn@doi [\apj] {10.1086/155518}, \href
  {https://ui.adsabs.harvard.edu/abs/1977ApJ...216..757D} {216, 757}

\bibitem[\protect\citeauthoryear{{Dearborn}, {Liebert}, {Aaronson}, {Dahn},
  {Harrington}, {Mould}  \& {Greenstein}}{{Dearborn}
  et~al.}{1986}]{Dearborn1986}
{Dearborn} D.~S.~P.,  {Liebert} J.,  {Aaronson} M.,  {Dahn} C.~C.,
  {Harrington} R.,  {Mould} J.,   {Greenstein} J.~L.,  1986, \mn@doi [\apj]
  {10.1086/163805}, \href
  {https://ui.adsabs.harvard.edu/abs/1986ApJ...300..314D} {300, 314}

\bibitem[\protect\citeauthoryear{{Downes} et~al.,}{{Downes}
  et~al.}{2004}]{Downes2004}
{Downes} R.~A.,  et~al., 2004, \mn@doi [\aj] {10.1086/383211}, \href
  {https://ui.adsabs.harvard.edu/abs/2004AJ....127.2838D} {127, 2838}

\bibitem[\protect\citeauthoryear{{Farihi}, {Arendt}, {Machado}  \&
  {Whitehouse}}{{Farihi} et~al.}{2018}]{Farihi2018}
{Farihi} J.,  {Arendt} A.~R.,  {Machado} H.~S.,   {Whitehouse} L.~J.,  2018,
  \mn@doi [\mnras] {10.1093/mnras/sty890}, \href
  {https://ui.adsabs.harvard.edu/abs/2018MNRAS.477.3801F} {477, 3801}

\bibitem[\protect\citeauthoryear{{Farihi}, {Sanders}, {Lilleengen},
  {Whitehouse}  \& {Erkal}}{{Farihi} et~al.}{2025}]{Farihi2025}
{Farihi} J.,  {Sanders} J.~L.,  {Lilleengen} S.,  {Whitehouse} L.~J.,   {Erkal}
  D.,  2025, \mn@doi [\mnras] {10.1093/mnras/staf1512}, \href
  {https://ui.adsabs.harvard.edu/abs/2025MNRAS.543..851F} {543, 851}

\bibitem[\protect\citeauthoryear{{Foreman-Mackey}, {Agol}, {Ambikasaran}  \&
  {Angus}}{{Foreman-Mackey} et~al.}{2017}]{Foreman-Mackey2017}
{Foreman-Mackey} D.,  {Agol} E.,  {Ambikasaran} S.,   {Angus} R.,  2017,
  \mn@doi [\aj] {10.3847/1538-3881/aa9332}, \href
  {https://ui.adsabs.harvard.edu/abs/2017AJ....154..220F} {154, 220}

\bibitem[\protect\citeauthoryear{{Frebel} \& {Norris}}{{Frebel} \&
  {Norris}}{2015}]{Frebel2015}
{Frebel} A.,  {Norris} J.~E.,  2015, \mn@doi [\araa]
  {10.1146/annurev-astro-082214-122423}, \href
  {https://ui.adsabs.harvard.edu/abs/2015ARA&A..53..631F} {53, 631}

\bibitem[\protect\citeauthoryear{{Gaia Collaboration} et~al.,}{{Gaia
  Collaboration} et~al.}{2023}]{Gaia2023}
{Gaia Collaboration} et~al., 2023, \mn@doi [\aap]
  {10.1051/0004-6361/202243940}, \href
  {https://ui.adsabs.harvard.edu/abs/2023A&A...674A...1G} {674, A1}

\bibitem[\protect\citeauthoryear{{Green}}{{Green}}{2013}]{Green2013}
{Green} P.,  2013, \mn@doi [\apj] {10.1088/0004-637X/765/1/12}, \href
  {https://ui.adsabs.harvard.edu/abs/2013ApJ...765...12G} {765, 12}

\bibitem[\protect\citeauthoryear{{Green} et~al.,}{{Green}
  et~al.}{2019}]{Green2019}
{Green} P.~J.,  et~al., 2019, \mn@doi [\apj] {10.3847/1538-4357/ab2bf4}, \href
  {https://ui.adsabs.harvard.edu/abs/2019ApJ...881...49G} {881, 49}

\bibitem[\protect\citeauthoryear{{Guidry} et~al.,}{{Guidry}
  et~al.}{2021}]{Guidry2021}
{Guidry} J.~A.,  et~al., 2021, \mn@doi [\apj] {10.3847/1538-4357/abee68}, \href
  {https://ui.adsabs.harvard.edu/abs/2021ApJ...912..125G} {912, 125}

\bibitem[\protect\citeauthoryear{{Gustafsson}, {Edvardsson}, {Eriksson},
  {J{\o}rgensen}, {Nordlund}  \& {Plez}}{{Gustafsson}
  et~al.}{2008}]{Gustafsson2008}
{Gustafsson} B.,  {Edvardsson} B.,  {Eriksson} K.,  {J{\o}rgensen} U.~G.,
  {Nordlund} {\r{A}}.,   {Plez} B.,  2008, \mn@doi [\aap]
  {10.1051/0004-6361:200809724}, \href
  {https://ui.adsabs.harvard.edu/abs/2008A&A...486..951G} {486, 951}

\bibitem[\protect\citeauthoryear{{Hansen}, {Andersen}, {Nordstr{\"o}m},
  {Beers}, {Placco}, {Yoon}  \& {Buchhave}}{{Hansen} et~al.}{2016}]{Hansen2016}
{Hansen} T.~T.,  {Andersen} J.,  {Nordstr{\"o}m} B.,  {Beers} T.~C.,  {Placco}
  V.~M.,  {Yoon} J.,   {Buchhave} L.~A.,  2016, \mn@doi [\aap]
  {10.1051/0004-6361/201527409}, \href
  {https://ui.adsabs.harvard.edu/abs/2016A&A...588A...3H} {588, A3}

\bibitem[\protect\citeauthoryear{{Harris} et~al.,}{{Harris}
  et~al.}{2018}]{Harris2018}
{Harris} H.~C.,  et~al., 2018, \mn@doi [\aj] {10.3847/1538-3881/aac100}, \href
  {https://ui.adsabs.harvard.edu/abs/2018AJ....155..252H} {155, 252}

\bibitem[\protect\citeauthoryear{{Heber}, {Bade}, {Jordan}  \& {Voges}}{{Heber}
  et~al.}{1993}]{Heber1993}
{Heber} U.,  {Bade} N.,  {Jordan} S.,   {Voges} W.,  1993, \aap, \href
  {https://ui.adsabs.harvard.edu/abs/1993A&A...267L..31H} {267, L31}

\bibitem[\protect\citeauthoryear{{Hippke}, {David}, {Mulders}  \&
  {Heller}}{{Hippke} et~al.}{2019}]{Hippke2019}
{Hippke} M.,  {David} T.~J.,  {Mulders} G.~D.,   {Heller} R.,  2019, \mn@doi
  [\aj] {10.3847/1538-3881/ab3984}, \href
  {https://ui.adsabs.harvard.edu/abs/2019AJ....158..143H} {158, 143}

\bibitem[\protect\citeauthoryear{{Ji}, {Cui}, {Liu}, {Luo}, {Zhao}  \&
  {Zhang}}{{Ji} et~al.}{2016}]{Ji2016}
{Ji} W.,  {Cui} W.,  {Liu} C.,  {Luo} A.,  {Zhao} G.,   {Zhang} B.,  2016,
  \mn@doi [\apjs] {10.3847/0067-0049/226/1/1}, \href
  {https://ui.adsabs.harvard.edu/abs/2016ApJS..226....1J} {226, 1}

\bibitem[\protect\citeauthoryear{{\DE{Kool}{De}{de}}~Kool \&
  {Green}}{{\DE{Kool}{De}{de}}~Kool \& {Green}}{1995}]{Kool1995}
{\DE{Kool}{De}{de}}~Kool M.,  {Green} P.~J.,  1995, \mn@doi [\apj]
  {10.1086/176051}, \href
  {https://ui.adsabs.harvard.edu/abs/1995ApJ...449..236D} {449, 236}

\bibitem[\protect\citeauthoryear{{Kraft}}{{Kraft}}{1967}]{Kraft1967}
{Kraft} R.~P.,  1967, \mn@doi [\apj] {10.1086/149359}, \href
  {https://ui.adsabs.harvard.edu/abs/1967ApJ...150..551K} {150, 551}

\bibitem[\protect\citeauthoryear{{Laos}, {Stassun}  \& {Mathieu}}{{Laos}
  et~al.}{2020}]{Laos2020}
{Laos} E.,  {Stassun} K.~G.,   {Mathieu} R.~D.,  2020, \mn@doi [\apj]
  {10.3847/1538-4357/abb3fe}, \href
  {https://ui.adsabs.harvard.edu/abs/2020ApJ...902..107L} {902, 107}

\bibitem[\protect\citeauthoryear{{Lenz} \& {Breger}}{{Lenz} \&
  {Breger}}{2005}]{Lenz2005}
{Lenz} P.,  {Breger} M.,  2005, \mn@doi [Communications in Asteroseismology]
  {10.1553/cia146s53}, \href
  {https://ui.adsabs.harvard.edu/abs/2005CoAst.146...53L} {146, 53}

\bibitem[\protect\citeauthoryear{{Li} et~al.,}{{Li} et~al.}{2018}]{Li2018}
{Li} Y.-B.,  et~al., 2018, \mn@doi [\apjs] {10.3847/1538-4365/aaa415}, \href
  {https://ui.adsabs.harvard.edu/abs/2018ApJS..234...31L} {234, 31}

\bibitem[\protect\citeauthoryear{{Liebert}, {Schmidt}, {Lesser}, {Stepanian},
  {Lipovetsky}, {Chaffe}, {Foltz}  \& {Bergeron}}{{Liebert}
  et~al.}{1994}]{Liebert1994}
{Liebert} J.,  {Schmidt} G.~D.,  {Lesser} M.,  {Stepanian} J.~A.,  {Lipovetsky}
  V.~A.,  {Chaffe} F.~H.,  {Foltz} C.~B.,   {Bergeron} P.,  1994, \mn@doi
  [\apj] {10.1086/173685}, \href
  {https://ui.adsabs.harvard.edu/abs/1994ApJ...421..733L} {421, 733}

\bibitem[\protect\citeauthoryear{{Margon}, {Kupfer}, {Burdge}, {Prince},
  {Kulkarni}  \& {Shupe}}{{Margon} et~al.}{2018}]{Margon2018}
{Margon} B.,  {Kupfer} T.,  {Burdge} K.,  {Prince} T.~A.,  {Kulkarni} S.~R.,
  {Shupe} D.~L.,  2018, \mn@doi [\apjl] {10.3847/2041-8213/aab42a}, \href
  {https://ui.adsabs.harvard.edu/abs/2018ApJ...856L...2M} {856, L2}

\bibitem[\protect\citeauthoryear{{Masci} et~al.,}{{Masci}
  et~al.}{2019}]{Masci2019}
{Masci} F.~J.,  et~al., 2019, \mn@doi [\pasp] {10.1088/1538-3873/aae8ac}, \href
  {https://ui.adsabs.harvard.edu/abs/2019PASP..131a8003M} {131, 018003}

\bibitem[\protect\citeauthoryear{{Moe} \& {Kratter}}{{Moe} \&
  {Kratter}}{2018}]{Moe2018}
{Moe} M.,  {Kratter} K.~M.,  2018, \mn@doi [\apj] {10.3847/1538-4357/aaa6d2},
  \href {https://ui.adsabs.harvard.edu/abs/2018ApJ...854...44M} {854, 44}

\bibitem[\protect\citeauthoryear{{Nebot G{\'o}mez-Mor{\'a}n} et~al.,}{{Nebot
  G{\'o}mez-Mor{\'a}n} et~al.}{2011}]{Nebot2011}
{Nebot G{\'o}mez-Mor{\'a}n} A.,  et~al., 2011, \mn@doi [\aap]
  {10.1051/0004-6361/201117514}, \href
  {https://ui.adsabs.harvard.edu/abs/2011A&A...536A..43N} {536, A43}

\bibitem[\protect\citeauthoryear{{Plant}, {Margon}, {Guhathakurta},
  {Cunningham}, {Toloba}  \& {Munn}}{{Plant} et~al.}{2016}]{Plant2016}
{Plant} K.~A.,  {Margon} B.,  {Guhathakurta} P.,  {Cunningham} E.~C.,  {Toloba}
  E.,   {Munn} J.~A.,  2016, \mn@doi [\apj] {10.3847/1538-4357/833/2/232},
  \href {https://ui.adsabs.harvard.edu/abs/2016ApJ...833..232P} {833, 232}

\bibitem[\protect\citeauthoryear{{Plez} \& {Cohen}}{{Plez} \&
  {Cohen}}{2005}]{Plez2005}
{Plez} B.,  {Cohen} J.~G.,  2005, \mn@doi [\aap] {10.1051/0004-6361:20042082},
  \href {https://ui.adsabs.harvard.edu/abs/2005A&A...434.1117P} {434, 1117}

\bibitem[\protect\citeauthoryear{{Rossi}, {Gigoyan}, {Avtandilyan}  \&
  {Sclavi}}{{Rossi} et~al.}{2011}]{Rossi2011}
{Rossi} C.,  {Gigoyan} K.~S.,  {Avtandilyan} M.~G.,   {Sclavi} S.,  2011,
  \mn@doi [\aap] {10.1051/0004-6361/201116704}, \href
  {https://ui.adsabs.harvard.edu/abs/2011A&A...532A..69R} {532, A69}

\bibitem[\protect\citeauthoryear{{Roulston}, {Green}, {Toonen}  \&
  {Hermes}}{{Roulston} et~al.}{2021}]{Roulston2021}
{Roulston} B.~R.,  {Green} P.~J.,  {Toonen} S.,   {Hermes} J.~J.,  2021,
  \mn@doi [\apj] {10.3847/1538-4357/ac157c}, \href
  {https://ui.adsabs.harvard.edu/abs/2021ApJ...922...33R} {922, 33}

\bibitem[\protect\citeauthoryear{{Shappee} \& {Thompson}}{{Shappee} \&
  {Thompson}}{2013}]{Shappee2013}
{Shappee} B.~J.,  {Thompson} T.~A.,  2013, \mn@doi [\apj]
  {10.1088/0004-637X/766/1/64}, \href
  {https://ui.adsabs.harvard.edu/abs/2013ApJ...766...64S} {766, 64}

\bibitem[\protect\citeauthoryear{{Shariat} \& {El-Badry}}{{Shariat} \&
  {El-Badry}}{2026}]{Shariat2026}
{Shariat} C.,  {El-Badry} K.,  2026, \mn@doi [\pasp]
  {10.1088/1538-3873/ae453b}, \href
  {https://ui.adsabs.harvard.edu/abs/2026PASP..138c4202S} {138, 034202}

\bibitem[\protect\citeauthoryear{{S{\"u}veges} et~al.,}{{S{\"u}veges}
  et~al.}{2015}]{Suveges2015}
{S{\"u}veges} M.,  et~al., 2015, \mn@doi [\mnras] {10.1093/mnras/stv719}, \href
  {https://ui.adsabs.harvard.edu/abs/2015MNRAS.450.2052S} {450, 2052}

\bibitem[\protect\citeauthoryear{{Tokovinin}, {Thomas}, {Sterzik}  \&
  {Udry}}{{Tokovinin} et~al.}{2006}]{Tokovinin2006}
{Tokovinin} A.,  {Thomas} S.,  {Sterzik} M.,   {Udry} S.,  2006, \mn@doi [\aap]
  {10.1051/0004-6361:20054427}, \href
  {https://ui.adsabs.harvard.edu/abs/2006A&A...450..681T} {450, 681}

\bibitem[\protect\citeauthoryear{{Van Eck} et~al.,}{{Van Eck}
  et~al.}{2017}]{VanEck2017}
{Van Eck} S.,  et~al., 2017, \mn@doi [\aap] {10.1051/0004-6361/201525886},
  \href {https://ui.adsabs.harvard.edu/abs/2017A&A...601A..10V} {601, A10}

\bibitem[\protect\citeauthoryear{{VanderPlas}}{{VanderPlas}}{2018}]{VanderPlas2018}
{VanderPlas} J.~T.,  2018, \mn@doi [\apjs] {10.3847/1538-4365/aab766}, \href
  {https://ui.adsabs.harvard.edu/abs/2018ApJS..236...16V} {236, 16}

\bibitem[\protect\citeauthoryear{{Vidotto} et~al.,}{{Vidotto}
  et~al.}{2014}]{Vidotto2014}
{Vidotto} A.~A.,  et~al., 2014, \mn@doi [\mnras] {10.1093/mnras/stu728}, \href
  {https://ui.adsabs.harvard.edu/abs/2014MNRAS.441.2361V} {441, 2361}

\bibitem[\protect\citeauthoryear{{Whitehouse}}{{Whitehouse}}{2023}]{Whitehouse2023}
{Whitehouse} L.,  2023, PhD thesis, University College London, UK

\bibitem[\protect\citeauthoryear{{Whitehouse}, {Farihi}, {Green}, {Wilson}  \&
  {Subasavage}}{{Whitehouse} et~al.}{2018}]{Whitehouse2018}
{Whitehouse} L.~J.,  {Farihi} J.,  {Green} P.~J.,  {Wilson} T.~G.,
  {Subasavage} J.~P.,  2018, \mn@doi [\mnras] {10.1093/mnras/sty1622}, \href
  {https://ui.adsabs.harvard.edu/abs/2018MNRAS.479.3873W} {479, 3873}

\bibitem[\protect\citeauthoryear{{Whitehouse}, {Farihi}, {Howarth}, {Mancino},
  {Walters}, {Swan}, {Wilson}  \& {Guo}}{{Whitehouse}
  et~al.}{2021}]{Whitehouse2021}
{Whitehouse} L.~J.,  {Farihi} J.,  {Howarth} I.~D.,  {Mancino} S.,  {Walters}
  N.,  {Swan} A.,  {Wilson} T.~G.,   {Guo} J.,  2021, \mn@doi [\mnras]
  {10.1093/mnras/stab1913}, \href
  {https://ui.adsabs.harvard.edu/abs/2021MNRAS.506.4877W} {506, 4877}

\bibitem[\protect\citeauthoryear{{Yoon} et~al.,}{{Yoon}
  et~al.}{2016}]{Yoon2016}
{Yoon} J.,  et~al., 2016, \mn@doi [\apj] {10.3847/0004-637X/833/1/20}, \href
  {https://ui.adsabs.harvard.edu/abs/2016ApJ...833...20Y} {833, 20}

\makeatother
\end{thebibliography}

\bsp	
\label{lastpage}
\end{document}